# Bidirectional spin-wave-driven domain wall motion in antiferromagnetically coupled ferrimagnets

Se-Hyeok Oh[1], Se Kwon Kim[2], Jiang Xiao[3,4,5], and Kyung-Jin Lee[1,6,7*]

[1]Department of Nano-Semiconductor and Engineering, Korea University, Seoul 02841, Korea

[2]Department of Physics and Astronomy, University of Missouri, Columbia, Missouri 65211, USA

[3]Department of Physics and State Key Laboratory of Surface Physics, Fudan University, Shanghai 200433, China

[4]Collaborative Innovation Center of Advanced Microstructures, Nanjing 210093, China

[5]Institute for Nanoelectronics Devices and Quantum Computing, Fudan University, Shanghai 200433, China

[6]Department of Materials Science and Engineering, Korea University, Seoul 02841, Korea

[7]KU-KIST Graduate School of Converging Science and Technology, Korea University, Seoul 02841, Korea

* corresponding email: kj_lee@korea.ac.kr

**Abstract**

**We investigate ferrimagnetic domain wall dynamics induced by circularly polarized spin waves theoretically and numerically. We find that the direction of domain wall motion depends on both the circular polarization of spin waves and the sign of net spin density of ferrimagnet. Below the angular momentum compensation point, left- (right-) circularly polarized spin waves push a domain wall towards (away from) the spin-wave source. Above the angular momentum compensation point, on the other hand, the direction of domain wall motion is reversed. This bidirectional motion originates from the fact that the sign of spin-wave-induced magnonic torque depends on the circular polarization and the subsequent response of the domain wall to the magnonic torque is governed by the net spin density. Our finding provides a way to utilize a spin wave as a versatile driving force for bidirectional domain wall motion.**

## I. Introduction

One class of ferrimagnets of emerging interest is a rare-earth (RE)-transition metal (TM) compound where the RE and TM moments are coupled antiferromagnetically. Owing to different Landé-g factors between the RE and TM elements, RE-TM ferrimagnets exhibit two unique compensation temperatures: the magnetic moment compensation temperature $T_M$ at which net magnetic moment vanishes, and the angular momentum compensation temperature $T_A$ at which net angular momentum vanishes [1-3].

Research on ferrimagnetic materials has focused on the understanding of their fundamental magnetism [4] and optical switching of magnetization [5-11]. Recently, RE-TM ferrimagnets attract renewed interest as they offer a material platform to investigate the antiferromagnetic spintronics [12-15]. Compared to ferromagnets that have served as core materials for spintronics research, antiferromagnets exhibit several distinct features such as the immunity to external field perturbations and fast dynamics due to antiferromagnetic exchange interaction. However, the external-field immunity of true antiferromagnets results in the experimental difficulty in both creating and controlling antiferromagnetic textures. On the other hand, RE-TM ferrimagnets have finite magnetic moment at the angular momentum compensation point at which the antiferromagnetic dynamics is realized. As a result, previously established creation and detection schemes for ferromagnets is directly applicable to RE-TM ferrimagnets. This simple but strong benefit of RE-TM ferrimagnets have recently initiated extensive studies on ferrimagnets, which include magnetization switching [16-19], domain wall motion [20-24], skyrmion (or bubble domain) motion [25-28], low damping [29], and efficient spin-transfer and spin-orbit torques due to antiferromagnetic alignment of atomic spins [30,31].

Among the previous studies listed above, the low damping of RE-TM ferrimagnets [29] is of particular interest from the view point of magnonic applications based on ferrimagnets because it enables a long-distance propagation of spin waves (SWs). For ferromagnets [32-39] and antiferromagnets [40-43], it was reported that a SW can move a DW by transferring its angular momentum or linear momentum. Though the SW property in ferrimagnets was established [44-47], the effect of SWs on ferrimagnetic domain wall (DW) motion remains unexplored. In comparison to ferromagnets and antiferromagnets, antiferromagnetically coupled ferrimagnets exhibit a distinguishing feature of SW eigenmodes. In ferromagnets, a spin wave (SW) with only one type of polarization is permitted, which drives a DW towards the SW source through the angular momentum transfer [34-37]. In antiferromagnets, however, both the left- and right-circularly polarized SWs are allowed and energetically degenerate, which can transfer the linear momentum to a DW through the SW reflection [40,41,43], resulting in the DW motion away from the SW source. In antiferromagnetically coupled ferrimagnets, on the other hand, the degeneracy of the two circularly polarized SWs can be lifted depending on the net spin density of ferrimagnet. Given that SW-induced DW motion in ferrimagnets has been unexplored, interesting and important questions remain unanswered: how a SW moves a ferrimagnetic DW and what the role of circular polarization of SW is.

In this paper, we study the dynamics of a ferrimagnetic DW induced by a SW in the vicinity of the angular momentum compensation temperature $T_A$. We investigate DW dynamics induced by left- and right-circularly polarized SWs [see Fig. 1(a) for an illustration of the two eigenmodes]. We begin with theoretical analysis based on the Lagrangian density and SW dispersion. We then conduct numerical simulation based on the atomistic Landau-Lifshitz-Gilbert (LLG) equation to confirm the analytical results. Our model system is shown in Fig. 1(b).

## II. Model

Our model system is a simple bipartite ferrimagnet which consists of two sublattices labeled by A and B. We introduce the staggered vector $\boldsymbol{n} = (\boldsymbol{A}_k - \boldsymbol{B}_k)/2$, and $\boldsymbol{m} = \boldsymbol{A}_k + \boldsymbol{B}_k$, where $\boldsymbol{A}_k$ and $\boldsymbol{B}_k$ are the unit vectors of spin moment at a site k that belongs to the sublattices A and B, respectively. The Lagrangian density for the ferrimagnet is given by [26,47-49]

$$\mathcal{L} = [-s\dot{\boldsymbol{n}} \cdot (\boldsymbol{n} \times \boldsymbol{m}) - \delta_s \mathbf{a}(\boldsymbol{n}) \cdot \dot{\boldsymbol{n}}] - \mathcal{U}, \tag{1}$$

where $s = (s_A + s_B)/2$, $\delta_s = s_A - s_B$, $s_i = M_i/\gamma_i$ is the angular momentum density, $M_i$ is the magnetic moment, $\gamma_i$ is the gyromagnetic ratio for sublattice $i$, $\mathbf{a}(\boldsymbol{n})$ is the vector potential for the magnetic monopole. The total energy $\mathcal{U}$ includes the exchange energy and anisotropy energy as

$$\mathcal{U} = \frac{a}{2}|\boldsymbol{m}|^2 + \frac{A}{2}(\nabla \boldsymbol{n})^2 - \frac{K}{2}(\hat{\boldsymbol{z}} \cdot \boldsymbol{n})^2 + \frac{\kappa}{2}(\hat{\boldsymbol{x}} \cdot \boldsymbol{n})^2, \tag{2}$$

where $a$ is the homogeneous exchange, $A$ is the inhomogeneous exchange, $K$ is the easy-axis anisotropy constant, and $\kappa$ is the hard-axis anisotropy constant. The Rayleigh function accounting for the dissipation is given by $\mathcal{R} = \alpha s \dot{\boldsymbol{n}}^2$ where $\alpha$ is the Gilbert damping constant. From the Lagrangian density and the Rayleigh dissipation, we obtain the equation of motion in terms of staggered vector $\boldsymbol{n}$ by integrating out the net magnetization variable $\boldsymbol{m}$ [26]:

$$\rho \boldsymbol{n} \times \ddot{\boldsymbol{n}} + 2\alpha s \boldsymbol{n} \times \dot{\boldsymbol{n}} + \delta_s \dot{\boldsymbol{n}} = \boldsymbol{n} \times \boldsymbol{f_n}, \tag{3}$$

where $\rho = s^2/a$ parametrizes the inertia and $\boldsymbol{f_n} = -\delta \mathcal{U}/\delta \boldsymbol{n}$ is the effective field. After linearizing the equations for small-amplitude fluctuations from the uniform state, we consider the SW ansatz as $\boldsymbol{n}(x,t) = \text{Re}\big[(n_x \exp[i(\omega t - kx)], n_y \exp[i(\omega t - kx)], 1)\big]$, where $n_x, n_y$ are the amplitudes of SW ($|n_x|, |n_y| \ll 1$), $\omega$ is the SW frequency, and $k$ is the wavevector. By solving the linearized equations with this ansatz, we obtain the dispersion

relation as

$$\omega_\pm = \frac{\pm\delta_s + \sqrt{\delta_s^2 + 4\rho(Ak^2 + K + \kappa/2)}}{2\rho}. \qquad (4).$$

Here the upper (lower) sign corresponds to the left- (right-) circularly polarized SW. The resonance frequencies for left- and right-circularly polarized SWs are different except at the angular momentum compensation point $T_A$ where the net spin density $\delta_s$ is zero. Figure 2 shows the agreement between the analytic SW dispersion relations [Eq. (4); lines] and numerical results that will be discussed below (symbols). Below or above $T_A$, the energy of right-circularly polarized SW differs from that of left-circularly polarized SW [see Fig. 2(a) and Fig. 2(c)]. At $T_A$ [Fig. 2(b)], two circularly polarized SWs are degenerate, which is analogous with antiferromagnetic SWs.

We next look into the dynamics of ferrimagnetic DW induced by SWs. We consider $\boldsymbol{n}$ as $\boldsymbol{n} = \boldsymbol{n_0} + \boldsymbol{\delta n}$ with DW texture $\boldsymbol{n_0}$ and small fluctuation $\boldsymbol{\delta n}$ ($|\boldsymbol{\delta n}| \ll |\boldsymbol{n_0}|$) with the constraint $\boldsymbol{n_0} \cdot \boldsymbol{\delta n} = 0$ to keep the unit length of $\boldsymbol{n}$ to linear order in $\boldsymbol{\delta n}$. We introduce two collective coordinates [50], the DW position $X(t)$ and center angle $\phi(t)$, and define a DW $\boldsymbol{n_0}$ by Walker ansatz [51], $\boldsymbol{n_0}(x,t) = (\sin\theta\sin\phi, \sin\theta\cos\phi, \cos\theta)$ where $\theta = 2\tan^{-1}[\exp\{(x-X)/\lambda\}]$ and $\lambda$ is the DW width. We consider the magnonic torque $\boldsymbol{\tau_m}$ that is given by [34,37,39,52]

$$\boldsymbol{\tau_m} = -A[(\boldsymbol{J_m} \cdot \boldsymbol{\nabla})\boldsymbol{n_0} - (\partial_x \rho_m)\boldsymbol{n_0} \times \partial_x \boldsymbol{n_0}], \qquad (5)$$

where magnon-flux density $J_m^x = \boldsymbol{n_0} \cdot \langle\boldsymbol{\delta n} \times \partial_x \boldsymbol{\delta n}\rangle$, and magnon number density $\rho_m = \langle\boldsymbol{\delta n}\rangle^2/2$. The first term in Eq. (5) represents the adiabatic magnonic torque rooted in the magnon current, and the second term represents the non-adiabatic magnonic torque caused by the gradient of the magnon density. Inserting Eq. (5) into the staggered LLG equation Eq. (3), we derive two coupled equations of motion as

$$M\ddot{X} - G\dot{\phi} + M\dot{X}/\tau = F_m, \qquad (6)$$

$$I\ddot{\phi} + G\dot{X} + I\dot{\phi}/\tau = -\kappa\lambda\sin 2\phi + T_m, \qquad (7)$$

where $M = 2\rho\mathcal{A}/\lambda$, $I = 2\rho\lambda\mathcal{A}$, $G = 2\delta_s\mathcal{A}$, and $\tau = \rho/\alpha s$ are the mass, the moment of inertia, the gyrotropic coefficient, and the relaxation time, respectively, and $\mathcal{A}$ is the cross sectional area of the DW. Here, $F_m = (2A/\lambda) \int dV [(\partial_x \rho_m) \boldsymbol{n_0} \times \partial_i \boldsymbol{n_0}]$ and $T_m = -2A \int dV [(\boldsymbol{J_m} \cdot \boldsymbol{\nabla}) \boldsymbol{n_0}]$ correspond to the magnon-induced force and torque, respectively. We note that the sign of $T_m$ is different for left- and right-circularly polarized SWs whereas the sign of $F_m$ is independent of the circular polarization of SW. It is because the sign of $J_m^x$ is different for left- and right-circularly polarized SWs whereas the sign of $\partial_x \rho_m$ is independent of the circular polarization of SW. From Eqs. (6) and (7), we finally obtain the steady-state velocity of DW below the Walker breakdown [53] as

$$v_{DW} = \frac{s}{2\mathcal{A}(\alpha^2 s^2 + \delta_s^2)} \left( \alpha \lambda F_m + \frac{\delta_s}{s} T_m \right), \tag{8}$$

which is the central result of this work. The first and second terms originate from non-adiabatic and adiabatic contributions, respectively. In Eq. (8), the ratio $\delta_s/s$ is an estimate of the degree how the dynamics of the system is close to that of ferromagnets. The condition of $\delta_s/2s \to \pm 1$ represents the ferromagnetic limit, whereas that of $\delta_s/2s \to 0$ represents the antiferromagnetic limit. In the ferromagnetic limit $(\delta_s/2s \to \pm 1)$, the second term in Eq. (8) becomes dominant for the DW motion. On the other hand, in the antiferromagnetic limit $(\delta_s/2s \to 0)$, the second term vanishes and the only first term is responsible for DW motion.

To verify Eq. (8), we perform micromagnetic simulations with the atomistic LLG equation. We start with the initial condition that the DW is located at the center of one-dimensional nanowire as shown in Fig. 1(b). SW is excited by an external AC field $\mathbf{B_{AC}}$ on the left side of DW. The atomistic LLG equation including the external AC field is given by

$$\frac{\partial \boldsymbol{S_i}}{\partial t} = -\gamma_i \boldsymbol{S_i} \times (\mathbf{B_{eff,i}} + \mathbf{B_{AC}}) + \alpha_i \boldsymbol{S_i} \times \frac{\partial \boldsymbol{S_i}}{\partial t}, \tag{9}$$

where $\boldsymbol{S_i}$ is the normalized spin moment vector, $\gamma_i = g_i \mu_B/\hbar$ is the gyromagnetic ratio, $\mu_B$ is the Bohr magneton, and $\alpha_i$ is the damping constant at a lattice site $i$. The odd (even) site $i$ corresponds to the TM (RE) element. $\mathbf{B_{eff,i}} = -\frac{1}{\mu_i} \frac{\partial \mathcal{H}}{\partial \boldsymbol{S_i}}$ is the effective field at each site, where $\mu_i$ is the magnetic moment per atom, one dimensional discrete Hamiltonian $\mathcal{H} = A_{sim} \sum_i \boldsymbol{S_i} \cdot \boldsymbol{S_{i+1}} - K_{sim} \sum_i (\boldsymbol{S_i} \cdot \hat{\boldsymbol{z}})^2 + \kappa_{sim} \sum_i (\boldsymbol{S_i} \cdot \hat{\boldsymbol{x}})^2$, $A_{sim}$ and $K_{sim}$ ($\kappa_{sim}$) are the exchange constant and easy (hard) axis anisotropy for simulations, respectively. To excite the SW, an external AC field $\mathbf{B_{AC}} = B_0[\cos\omega t\,\hat{x} \pm \sin\omega t\,\hat{y}]$ is applied on two cells at 252 nm away from the DW. We use the following simulation parameters: $A_{sim} = 1.64$ meV, $K_{sim} = 6.47$ μeV, $\kappa_{sim} = 0.02 K_{sim}$, $B_0 = 100$ mT, the lattice constant is 0.42 nm, and the Landé $g$–factor $g_{RE} = 2$ for rare earth and $g_{TM} = 2.2$ for transition metal [54]. We consider the damping constant is uniform for all sites, i.e., $\alpha_{RE} = \alpha_{TM} = 5 \times 10^{-4}$ for simplicity. We use magnetic moments $M_{RE}$ and $M_{TM}$ as listed in TABLE 1.

### III. Results and Discussion

Figure 3(a)-(c) show the simulation results of DW velocity as a function of the SW frequency. Figure 3(a) represents the results for the case below the angular momentum compensation point $T_A$. As the SW gap is different for left- and right-circularly polarized SWs [Fig. 2(a)], the threshold SW frequency for the DW motion is also different for left- and right-circularly polarized SWs. An interesting observation for the DW motion is that the moving direction of DW depends on the circular polarization of SW. Left- (Right-) circularly polarized SW moves the DW towards (away from) the SW source. This bi-directional DW motion is understood by the fact that left- and right-circularly polarized SWs carry the angular momentum with opposite signs. When the SW passes through the DW, the angular momentum of SW is transferred to the DW so that the DW moving direction depends on the circular polarization of SW. This is directly related to the fact that the sign of $T_m$ is different for left- and right-circularly polarized SWs, whereas the sign of $F_m$ is independent of the circular polarization of SW. Given that $T_m$ and $F_m$ in Eq. (8) respectively correspond to contributions from the adiabatic and non-adiabatic magnonic torques, the bi-directional DW motion depending on the circular polarization of SW evidences that the adiabatic magnonic torque is dominant over the non-adiabatic one.

Solid and dashed lines in Fig. 3(a) are calculated from Eq. (8), with $F_m$ and $T_m$ obtained from numerical calculations. We find that the numerically obtained bi-directional behavior (symbols) is reasonably described by Eq. (8) in high-frequency ranges. In low-frequency ranges, a discrepancy between Eq. (8) and numerical results appears possibly because of nonlinear effects, which are not captured by our current analytical models. The results for the case above the angular momentum compensation point $T_A$ [Fig. 3(c)] can be understood in a similar way. Contrary to the case below $T_A$, overall spin moments in the system are reversed so that left- (right-) circularly polarized SW makes DW move away from (towards) the source.

To further elucidate SW-induced ferrimagnetic DW motion below and above $T_A$, we investigate the spin current $J_S$, which is defined as $\boldsymbol{J_s} = -A\langle \boldsymbol{n} \times \partial_x \boldsymbol{n}\rangle$. Figure 4 shows the schematic of SW transmission through a DW (top panel) and the $z$ component of the spin current $J_S^z$ along the propagation direction (i.e., the $x$ axis, bottom panel). For the left-circularly polarized SW (solid line), the spin current in the left domain part decreases gradually due to the damping. After the SW passes through the DW, the spin current abruptly flips its sign due to overall reversal of spin moments. The spin-current change is transferred to the DW, resulting in the DW motion. For the right-circularly polarized SW (dashed line), overall sign of the spin current is reversed. It is the reason that the direction of DW propagation is the opposite for left- and right-circularly polarized SWs. The sign of the spin-current change is the same below and above $T_A$, but $\delta_s$ changes its sign [see Eq. (8)] because the spin directions in the domain part changes accordingly, which results in the sign difference of the DW velocity below and above $T_A$.

For the case at the angular momentum compensation point $T_A$ (i.e., $\delta_s = 0$), both left- and right-circularly polarized SWs drive the DW to the same direction (i.e., towards the SW source) as shown in Fig. 3(b). We note that this DW moving direction at $T_A$ is the opposite to the direction of the DW motion induced by circularly polarized spin waves in true antiferromagnets [40,41]. In true antiferromagnets where the shape anisotropy is absent, circularly polarized SWs make the DW precess, which results in the SW reflection. The reflected SWs transfer linear momentum to DW, and push the DW away from the SW source. For the case at $T_A$ of ferrimagnets, however, the net magnetic moment is finite so that the shape anisotropy does not vanish. As a result, the DW experiences the hard-axis anisotropy, which prevents the DW precession. Therefore, the ferrimagnetic DW still serves as a reflectionless potential called the Pöschl-Teller potential [55] and its motion is governed by the force from the magnonic torque [$\alpha\lambda F_m$ term in Eq. (8)]. As the sign of $F_m$ is independent of the SW circular polarization, both left- and right-circularly polarized SWs pull the DW along the same direction, i.e., towards the SW source. This force-induced motion of the ferrimagnetic DW toward the spin-wave source at $T_A$ is similar to the motion of the antiferromagnetic DW toward the spin-wave source for linearly-polarized spin waves reported in Ref. [40].

Dependence of the DW velocity on the SW circular polarization is summarized in Fig. 3(d), which shows the DW velocity as a function of the net spin density $\delta_s$ at a fixed SW frequency ($\omega$ = 0.7 THz). The sign of DW velocity depends not only on the circular polarization, but also on the sign of the net spin density. We note that the DW velocity is not zero at $T_A$ because the adiabatic and non-adiabatic contributions are not compensated at $T_A$.

### IV. Summary

We have investigated the SW circular-polarity dependence of ferrimagnetic DW dynamics theoretically and numerically. We find that the DW moves along the opposite direction depending on the circular polarization of SW. This bi-directional DW motion is caused by the fact that the signs of the spin current and the angular momentum transferred to DW are opposite for left- and right-circularly polarized SWs. The overall tendency of DW moving direction is reversed when the sign of the net spin density $\delta_s$ is reversed. At $T_A$ where the angular momentum vanishes, the dissipative non-adiabatic magnonic torque is the main driving force so that DW moves along the same direction (towards the SW source) regardless of the SW circular polarization.

Our finding of bi-directional ferrimagnetic DW driven by SWs can be generalized to other ferrimagnetic topological excitations such as magnetic skyrmions and vortices. This bi-directionality of ferrimagnetic DW motion depending either on the SW circular polarization or on the sign of the net spin density will be useful for magnonic spintronics [56] because such bi-directional motion, which makes the device functionality versatile, can be realized without moving the location of a SW source.

**Acknowledgement**

This work was supported by the National Research Foundation of Korea (NRF) (Grants No. 2015M3D1A1070465 and No. 2017R1A2B2006119), the KIST Institutional Program (Project No. 2V05750). S.K.K. was supported by the startup fund at the University of Missouri. J.X. was supported by the National Natural Science Foundation of China (Grants No. 11722430).

**Table 1.** Used magnetic moments $M_{TM}$ and $M_{RE}$ for transition metal and rare earth elements, respectively, in simulation. Index 5 coincides with the angular momentum compensation point $T_A$.

| Index | 1 | 2 | 3 | 4 | 5 | 6 | 7 | 8 | 9 |
|---|---|---|---|---|---|---|---|---|---|
| $M_{TM}$(kA/m) | 460 | 455 | 450 | 445 | 440 | 435 | 430 | 425 | 420 |
| $M_{RE}$(kA/m) | 440 | 430 | 420 | 410 | 400 | 390 | 380 | 370 | 360 |

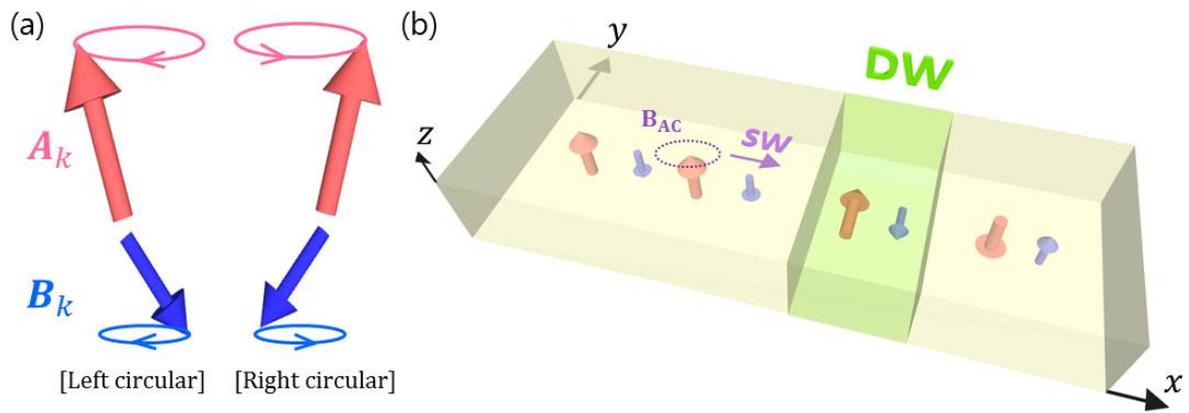

**Figure 1.** (a) Illustration of left- and right-circularly spin waves in an antiferromagnetically coupled ferrimagnet. (b) Schematic graphic of one-dimensional ferrimagnetic nanowire with a domain wall (DW). Domain wall is positioned at the center of nanowire. Spin wave is excited by an external AC field ($\mathbf{B}_{AC}$) on the left side (252 nm apart from DW).

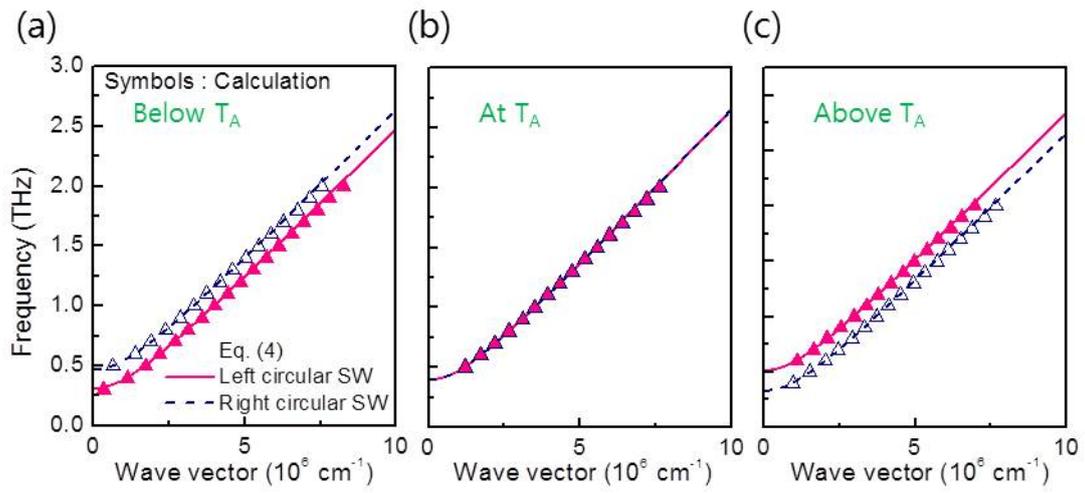

**Figure 2.** Spin-wave dispersion relations (a) below $T_A$, (b) at $T_A$, and (c) above $T_A$. Symbols represent numerical simulation results and lines represent Eq. (4). Solid (Open) triangular symbols correspond to left- (right-) circularly polarized spin wave.

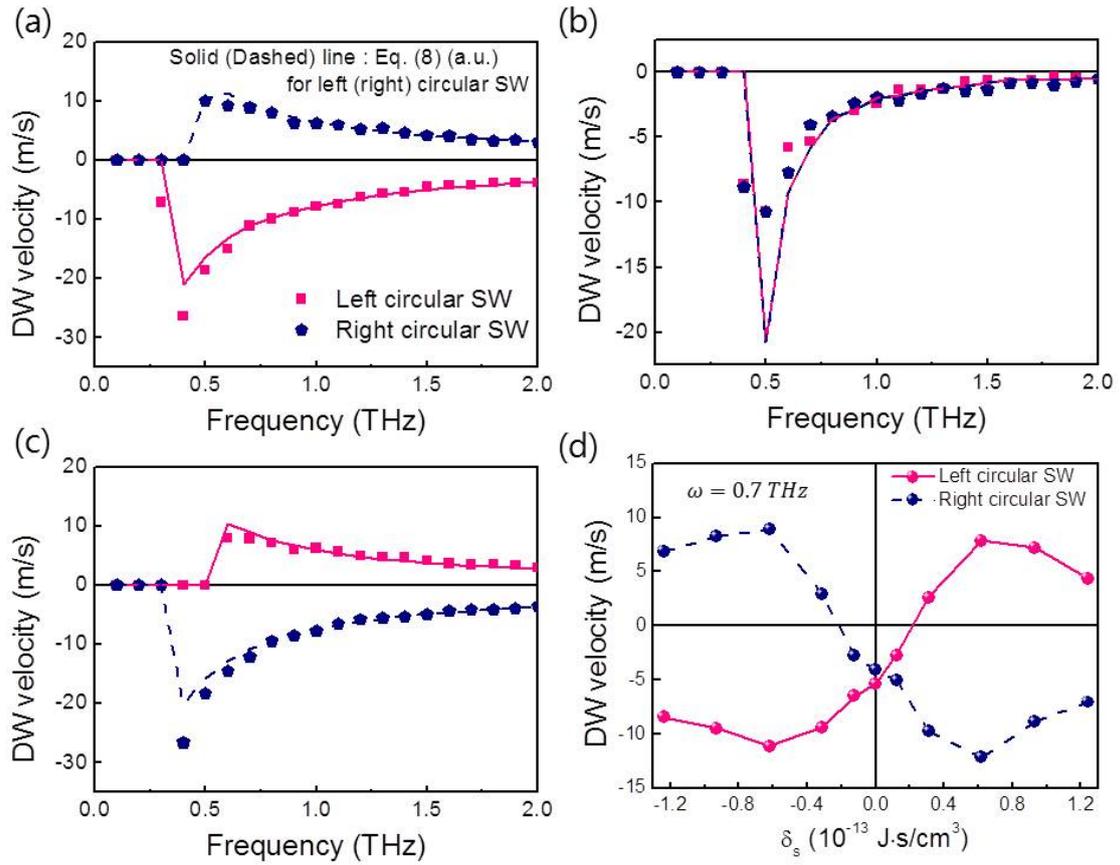

**Figure 3.** Calculated domain wall velocity results (a) below $T_A$, (b) at $T_A$, and (c) above $T_A$ with various spin-wave frequencies. Symbols are the simulation results and lines are Eq. (8) (arbitrary unit). (d) Domain wall velocity as a function of the net spin density $\delta_s$ at a fixed frequency $\omega = 0.7$ THz. Negative $\delta_s$ corresponds to the case below $T_A$.

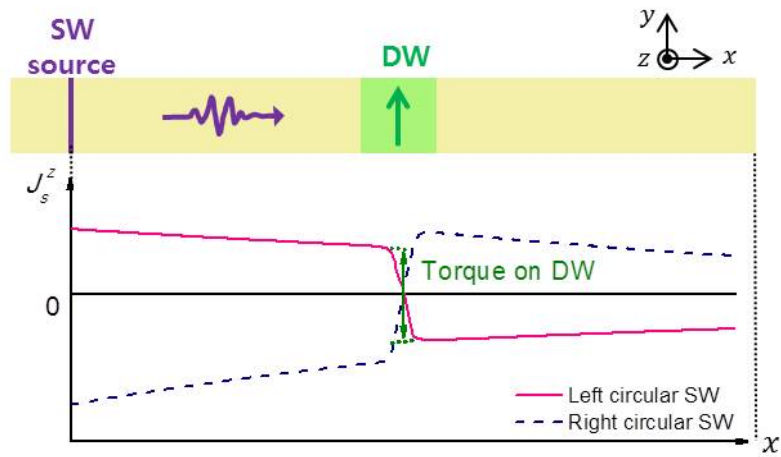

**Figure 4.** Schematic of transmitted SW (top) and the $z$ component of the spin current $J_s$ along the wire length. A DW is positioned at the atomic site $i = 2000$, and SW source is at $i = 1700$. Assumed parameters are those with the index 3 (i.e., below $T_A$) listed in the TABLE 1 and the SW frequency is 0.6 THz.

**Reference**

1. R. Wangness, Phys. Rev. **91**, 1085 (1953).

2. M. Binder *et al*., Phys. Rev. B **74**, 134404 (2006).

3. C. Stanciu, A. Kimel, F. Hansteen, A. Tsukamoto, A. Itoh, A. Kirilyuk, and T. Rasing, Phys. Rev. B **73**, 220402 (2006).

4. J. M. D. Coey, "Magnetism and Magnetic Materials", Cambridge University Press (2010).

5. C. D. Stanciu, F. Hansteen, A. V. Kimel, A. Tsukamoto, A. Itoh, and T. Rasing, Phys. Rev. Lett. **99**, 047601 (2007).

6. I. Radu *et al*., Nature **472**, 205 (2011).

7. S. Mangin *et al*., Nat. Mater. **13**, 286 (2014).

8. W. Cheng, X. Li, H. Wang, X. Cheng, and X. Miao, AIP Advances **7**, 056018 (2017).

9. S. Alebrand *et al*., Appl. Phys. Lett. **101**, 162408 (2012).

10. K. Vahaplar *et al*., Phys. Rev. B **85**, 104402 (2012).

11. T. A. Ostler *et al*., Nat. Commun. **3**, 666 (2012).

12. T. Jungwirth, X. Marti, P. Wadley, and J. Wunderlich, Nat. Nanotechnol. **11**, 231 (2016).

13. T. Jungwirth, J. Sinova, A. Manchon, X. Marti, J. Wunderlich, and C. Felser, Nat. Phys. **14**, 200 (2018).

14. O. Gomonay, V. Baltz, A. Brataas, and Y. Tserkovnyak, Nat. Phys. **14**, 213 (2018).

15. R. A. Duine, K.-J. Lee, S. S. P. Parkin, and M. D. Stiles, Nat. Phys. **14**, 217 (2018).

16. N. Roschewsky *et al*., Appl. Phys. Lett. **109**, 112403 (2016).

17. J. Finley and L. Liu, Phys. Rev. Applied **6**, 054001 (2006).

18. K. Ueda, M. Mann, C.-F. Pai, A.-J. Tan, and G. S. D. Beach, Appl. Phys. Lett. **109**, 232403 (2016).

19. R. Mischra, J. Yu, X. Qiu, M. Motapothula, T. Venkatesan, and H. Yang, Phys. Rev. Lett. **118**, 167201 (2017).

20. K.-J. Kim *et al.*, Nat. Mater. **16**, 1187 (2017).

21. S.-H. Oh, S. K. Kim, D.-K. Lee, G. Go, K.-J. Kim, T. Ono, Y. Tserkovnyak, and K.-J. Lee, Phys. Rev. B **96**, 10047(R) (2017).

22. L. Caretta *et al.*, Nat. Nanotechnol. **13**, 1154 (2018).

23. S. A. Siddiqui, J. Han, J. T. Finley, C. A. Ross, and L. Liu, Phys. Rev. Lett. **121**, 057701 (2018).

24. S.-H. Oh and K.-J. Lee, J. Magn. **23**, 196 (2018).

25. M. Tanaka, H. Kanazawa, S. Sumitomo, S. Honda, K. Mibu, and H. Awano, Appl. Phys. Express **8**, 073002 (2015).

26. S. K. Kim, K.-J. Lee, and Y. Tserkovnyak, Phys. Rev. B **95**, 140404(R) (2017).

27. S. Woo *et al.*, Nat. Commun. **9**, 959 (2018).

28. Y. Hirata *et al.*, Nat. Nanotechnol. **14**, 232 (2019).

29. D.-H. Kim *et al.*, Phys. Rev. Lett. **122**, 127203 (2019).

30. J. Yu *et al.*, Nat. Mater. **18**, 29 (2019).

31. T. Okuno *et al.*, arXiv:1903.03251 (2019).

32. D.-S. Han, S.-K. Kim, J.-Y. Lee, S. J. Hermsdoerfer, H. Schultheiss, B. Leven, and B. Hillebrands, Appl. Phys. Lett. **94**, 112502 (2009).

33. S.-M. Seo, H.-W. Lee, H. Kohno, and K.-J. Lee, Appl. Phys. Lett. **98**, 012514 (2011).

34. P. Yan, X. Wang, and X. Wang, Phys. Rev. Lett. **107**, 177207 (2011).

35. X.-G. Wang, G.-H. Guo, Y.-Z. Nie, G.-F. Zhang, and Z.-X. Li, Phys. Rev. B **86**, 054445 (2012).

36. P. Yan, A. Kamra, Y. Cao, and G. E. W. Bauer, Phys. Rev. B **88**, 144413 (2013).

37. S. K. Kim and Y. Tserkovnyak, Phys. Rev. B **92**, 020410 (2015).

38. W. Wang *et al.*, Phys. Rev. Lett. **114**, 087203 (2015).

39. K.-W. Kim *et al.*, Phys. Rev. Lett. **122**, 147202 (2019).

40. E. G. Tveten, A. Qaiumzadeh, and A. Brataas, Phys. Rev. Lett. **112**, 147204 (2014).

41. S. K. Kim, Y. Tserkovnyak, and O. Tchernyshyov, Phys. Rev. B **90**, 104406 (2014).

42. J. Lan, W. Yu, and J. Xiao, Nat. Commun. **8**, 178 (2017).

43. W. Yu, J. Lan, and J. Xiao, Phys. Rev. B **98**, 144422 (2018).

44. S. Pikin, Sov. Phys. JETP **27**, 995 (1968).

45. A. Andreev and V. Marchenki, Zh. Eksp. Teor, Fiz. **70**, 1522 (1976).

46. E. Solano-Carrillo, R. Franco, and J. Silva-Valencia, Solid State Commun. **150**, 2061 (2010).

47. A. F. Andreev and V. I. Marchenko, Soviet Physics Uspekhi **23**, 21 (1980).

48. A. Chiolero and D. Loss, Phys. Rev. B **56**, 738 (1997).

49. B. A. Ivanov and A. L. Sukstanskii, Solid State Commun. **50**, 523 (1984).

50. O. Tretiakov, D. Clarke, G.-W. Chern, Y. B. Bazaliy, and O. Tchernyshyov, Phys. Rev. Lett. **100**, 127204 (2008).

51. L. D. Landau, Course of Theoretical Physics **8**, 15 (1960).

52. R. Khoshlahni, A. Qaiumzadeh, A. Bergman, and A. Brataas, Phys. Rev. B **99**, 054423 (2018).

53. N. L. Schryer and L. R. Walker, J. Appl. Phys. **45**, 5406 (1974).

54. J. Jensen and A. R. Mackintosh, Rare Earth Magnetism (Clarendon, Oxford, UK,

1991).

55. G. Pöschl and E. Teller, Zeitschrift für Physik **83**, 143 (1933).

56. A. V. Chumak, V. I. Vasyuchka, A. A. Serga, and B. Hillebrands, Nat. Phys. **11**, 453 (2015).